\documentclass[12pt,a4paper]{article}
\usepackage{graphicx}
\usepackage{amsmath}
\usepackage{amssymb}
\usepackage[unicode,colorlinks=false]{hyperref}

\graphicspath{{fig/}}

\begin{document}

\title{Asymmetric double well system as effective model for the kicked one}
\author{V.I. Kuvshinov\thanks{E-mail:V.Kuvshinov@sosny.bas-net.by}, A.V.
Kuzmin\thanks{E-mail:avkuzmin@sosny.bas-net.by} and V.A.
Piatrou\thanks{E-mail:vadzim.piatrou@gmail.com}\\
{\small Joint Institute for Power and Nuclear Research,} \\ {\small Krasina str.
99, Minsk,  220109, Belarus }}
\date{}
\maketitle

\begin{abstract}
Effective Hamiltonian for the kicked double well system was derived using the
{Campbell}-{Baker}-{Hausdorff} expansion formula. Asymmetric model for the
kicked system was constructed. Analytical description of the quasienergy levels
splittings for the low laying doublets was given in the framework of the model.
Numerical calculations confirm applicability of the proposed effective
asymmetric approach for the double well system with the kick-type perturbation.
\end{abstract}

\section{Introduction}

The connection between the semiclassical properties of chaotic systems and
purely quantum processes such as tunneling is a reach rapidly developing field
of research nowadays. Our insight in some novel phenomena in this field was
extended in the last decades. The most intriguing among them are the chaos
assisted tunneling (CAT) and the closely related coherent destruction of
tunneling (CDT).

The first one is an enhancement of tunneling in perturbed low-dimensional
systems at relatively high external field strengths and high driving frequencies
(in order the singlet-doublet crossings to occur)~\cite{Lin:90, Plata:92,
Holthaus:92}. This phenomenon takes place when levels of the regular doublet
undergo an avoided crossing with the chaotic state \cite{Bohigas:93,Latka:94}.
At the semiclassical level of description one considers tunneling between
KAM-tori embedded into the "chaotic sea". The region of chaotic motion affects
tunneling rate because compared to direct tunneling between tori it is easier
for the system to penetrate primarily into the chaotic region, to travel then
along some classically allowed path and finally to tunnel onto another
KAM-torus~\cite{Utermann:94,Mouchet:01}.

CDT phenomenon is a suppression of tunneling when values of amplitude and 
frequency of driving force belong to some one-dimensional manifold  in the
perturbation parameters' space~\cite{Grossmann:91}. This phenomenon occurs due
to the exact crossing of two  states with different symmetries from the
tunneling doublet. In this parameter region tunneling time diverges which means
the total localization of quantum state on the initial torus.

CAT phenomenon as well as CDT were experimentally observed in a number of real
physical systems. The CAT observation  between whispering gallery-type modes of
microwave cavity having the form of the annular billiard was reported in the
Ref.~\cite{Dembowski:00}. The same phenomenon for ultracold atoms was
experimentally investigated in Refs.~\cite{Steck:01,Hensinger:01}. The study of
the dielectric microcavities provided evidences for CAT as
well~\cite{Podolskiy:05}. Both phenomena were observed in two coupled optical
waveguides \cite{vorobeichik:03,Valle:07}. Recently experimental evidence of
coherent control of single particle tunneling in strongly driven double well
potential was reported in Ref. \cite{kierig:08}.

The most common methods which are used to investigate the CAT are numerical
methods based on the Floquet theory~\cite{Utermann:94,Shirley:65,Grifoni:98}.
Among other approaches to CAT we would like to mention the scattering approach
for billiard systems~\cite{Frischat:98,Doron:95} and quantum mechanical
amplitudes in the complex configuration space~\cite{Shudo:96,Shudo:98,Shudo:01}.
There is an analytical approach based on instanton technique, which was proposed
in \cite{KK:05, KKS:03, KKS:02} and independently used in~\cite{Igarashi:06}.
Alternative approach based on quantum instantons which are defined using an
introduced notion of quantum action was suggested in~\cite{Jirari:01}.

We will investigate a quasienergy spectrum of paradigmatic model for different
physical systems, namely double well potential. We consider this system with
perturbation of the kicked type. One of the most attractive features of kicked
systems is the well investigated simple quantum map which stroboscopically
evolves the system from kick $n$ to kick $n+1$ and greatly facilitates
theoretical analysis.

The main idea of this paper is to study the possibility to construct an
effective autonomous model for the non-autonomous perturbed system using
{Campbell}-{Baker}-{Hausdorff} expansion formula and to test it in numerical
calculations of the quasienergy spectrum. Both CAT and CDT are connected with
the behavior of the quasienergy spectrum (avoided or exact crossing of the
levels). Thus the development of the methods for calculation of this spectrum is
important for extending one's knowledge in CAT and CDT. We regard the kick-type
perturbation which is proportional to $x$. This perturbation, in contrast to
perturbation proportional to $x^2$, destroys the spacial symmetry in the system
which is important  for presence of the CAT phenomenon~\cite{Peres:91}. The main
role in quantum dynamics of our system is played by the classical asymmetry.
There is no chaos induced processes in it but in our future work we will use
this approach to system with CAT and CDT.

In this paper we propose the effective model for the kicked double well system
which gives a possibility to simplify the numerical calculations of the
quasienergy spectrum and allows to determine both analytically and numerically
the quasienergy splitting dependence on both the perturbation strength and
frequency.

\section{Effective Hamiltonian for the kicked system}

Now lets construct the effective Hamiltonian for the double well system with the
perturbation of the kick-type. Hamiltonian of the particle in the double-well
potential can be written in the following form:
\begin{equation}\label{eq:H}
H_0 = \frac{p^2}{2 m} + a_0\, x^4 - a_2\, x^2,
\end{equation}
where $m$ - mass of the particle, $a_0, a_2$ - parameters of the
potential.

We consider the perturbation of the kick-type which is proportional to $x$
\begin{equation}\label{eq:V}
V_{per} = \epsilon\, x \sum^{+ \infty}_{n = - \infty} \delta(t- n
T),
\end{equation}
where $\epsilon$ - perturbation strength, $T$ -  perturbation period, $t$ -
time.

Full Hamiltonian of the system is the following:
\begin{equation}\label{SystemHamiltonian}
H = H_0 + V_{per}.
\end{equation}

Now we will construct an effective Hamiltonian for the system under
investigation using the following definition:
\begin{equation}\label{def:eff}
	exp(- i H_{eff} T) = exp(- i \epsilon x) exp(- i H_0 T),
\end{equation} 
where RHS is a one-period evolution operator. We restrict our consideration by
sufficiently small values of both the perturbation strength and period. Using
the {Campbell}-{Baker}-{Hausdorff} expansion formula for the kicked dynamical
systems we can rewrite the last expression (\ref{def:eff}) in the following way
\cite{scharf:88}:
\begin{equation}\label{He1}
H_{eff} = H_0 + \frac{\epsilon \, \nu}{2 \pi} \int^1_0 ds \, g\left[exp(- i
\epsilon s \hat{x}) exp(- i \hat{H}_0 T)\right] x,
\end{equation} 
where
\[
g(z) = \frac{ln\,z}{z-1} = \sum^\infty_{n=0} \frac{(-1)^n}{n + 1} (z-1)^n \qquad
\text{and} \qquad \nu = \frac{2 \pi}{T}.                                        
                    \]
With the definition of $g(z)$ formula (\ref{He1}) can be expanded in the
following form:
\begin{equation}\label{He2}
	H_{eff} = H_0 + \frac{\epsilon \, \nu}{2 \pi} \sum^\infty_{n=0}
\frac{1}{n + 1} \sum^n_{k=0} (-1)^k \frac{n!}{k! (n-k)!} \int^1_0 ds \,
\left[exp(- i \epsilon s \hat{x}) exp(- i \hat{H}_0 T)\right]^k x,
\end{equation} 
where the expression under the integral is a map in the power $k$ which for
sufficiently small values of both the perturbation strength and period can be
rewritten as follows:
\begin{equation}\label{map}
	\left[exp(- i \epsilon s \hat{x}) exp(- i \hat{H}_0 T)\right]^k x = x -
\frac{k T}{m} p + O(\epsilon^2, \epsilon T, T^2).
\end{equation} 
Substituting (\ref{map})  in expression (\ref{He2}) one obtains the following
form of the effective Hamiltonian:
\[H_{eff} = \frac{p^2}{2 m} + a_0 x^4 - a_2 x^2 + \frac{\epsilon \, \nu}{2 \pi}
x + \frac{\epsilon}{2 m} p + O(\epsilon^2, \epsilon T, T^2).\]
The fouth term in RHS of the last expression has the same order in perturbation
parameters as first three main terms. The fifth term is proportional to small
parameter, namely perturbation strength. We restrict our consideration by terms
without small parameters and neglect all terms with order higher than zero in
the perturbations parameters. As a result we have the following effective
Hamiltonian for the kicked double well system:
\begin{equation}\label{def:hameff}
H_{eff} = \frac{p^2}{2 m} + a_0 x^4 - a_2 x^2 + \frac{\epsilon \, \nu}{2 \pi} x.
\end{equation} 
This is the Hamiltonian for the asymmetric double well potential without
perturbation. In contrast to the kicked system it is autonomous. In the next
section we will consider the properties for this system and construct the
effective asymmetric model for the kicked system. In section \ref{sec:num} the
correspondence between kicked and asymmetric effective double well systems will
be tested numerically.

\section{Effective asymmetric model for the kicked system}\label{sec:twosys}

\begin{figure}[!ht]
\centering
\includegraphics[angle = 270,width=0.6\textwidth]{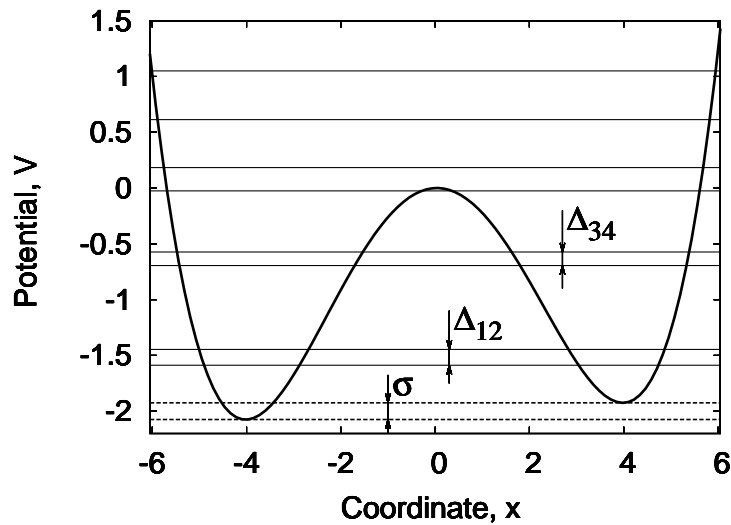}
\caption{Asymmetric double well potential (thick solid line) with eight lowest
energy levels (thin solid lines). Minima shift  ($\sigma = 0.15$) is shown by
dashed lines. The model parameters are $m~=~1,$ $a_0~=~1/128,$ $a_2 =
1/4$.}\label{fig:adwp}
\end{figure}

Hamiltonian of the classical particle in the asymmetric double-well potential is
the following:
\begin{equation}\label{def:hamas}
H_{as} = \frac{p^2}{2 m} + a_0 x^4 - a_2 x^2 + \sigma \sqrt{\frac{a_0}{2 a_2}}
x,                                                     \end{equation} 
where $a_0$ and $a_2$ are parameters of the potential, $\sigma$ - asymmetric
parameter. The asymmetric potential with parameters $m~=~1,$ $a_0~=~1/128,$ $a_2
= 1/4$ and $\sigma = 0.15$ is shown in the figure \ref{fig:adwp} (thick solid
line). Eight lowest energy levels are shown in the same figure with thin solid
lines. The form of the last term in the RHS of the Hamiltonian 
(\ref{def:hamas}) is more handy due to special choice of the asymmetric
parameter $\sigma$. Parameter $\sigma$ for this form of the Hamiltonian is equal
to shift between bottoms of the wells (dashed lines in the
figure~\ref{fig:adwp}).

In order to construct the effective model for the kicked system we should to
introduce an effective value of the shift in the kicked system. Comparing
obtained effective Hamiltonian~(\ref{def:hameff}) and Hamiltonian for the
asymmetric double well system (\ref{def:hamas}) we define effective asymmetric
parameter in the following way:
\begin{equation}\label{def:sigma}
\sigma_{eff} = \sqrt{\frac{a_2}{2 a_0}} \frac{\epsilon \, \nu}{\pi}.
\end{equation} 

Now we can give definition of the proposed effective model: asymmetric double
well system with the effective parameter $\sigma_{eff}$ defining by expression
(\ref{def:sigma}) is an effective model for the kicked system with the
perturbation parameters $\epsilon$ and $\nu$. The parameters $a_0$, $a_2$  and
$m$ are the same for both systems.

The definition of the effective asymmetric parameter (\ref{def:sigma}) shows
that perturbation strength and frequency appears in it as a product $\epsilon
\nu$. This is the first advantage of the effective approach. We have effectively
only one asymmetric parameter ($\sigma$) instead of two perturbation parameters
($\epsilon$ and $\nu$). The second advantage is the more simple way of the
numerical calculations which will be discussed in the next section.

The third advantage of the proposed approach is that splitting for the doublets
laying below the barrier hump in the asymmetric double well can be described
analytically. The asymmetric model can be considered as a pair of shifted
harmonic oscillators. The shift is equal to asymmetric parameter $\sigma$. It is
obvious that in this constructed system non-degenerate energy doublets have
splitting $\Delta = \sigma$. In asymmetric system splitting between levels
remains close to $\sigma$ for all doublets lying below the barrier top as for
case shown in the figure~\ref{fig:adwp}. This correspondence can be used in
order to give analytical description of the quasienergy spectrum in the kicked
system. Using expression (\ref{def:sigma}) we can write down the formula for low
laying quasienergy doublets' splittings of the time-dependent system
(\ref{SystemHamiltonian})
\begin{equation}\label{an-fomula}
\Delta = \sqrt{\frac{a_2}{2 a_0}} \frac{\epsilon \, \nu}{\pi}.
\end{equation} 
It worth to mention the linear dependence of the levels splitting on both the
perturbation strength and frequency. The applicability of this analytical
description will be tested in the numerical calculations in the next section.

\section{Numerical calculations}\label{sec:num}

For the computational purposes it is convenient to choose the eigenvectors of
harmonic oscillator as basis vectors. In this
representation matrix elements of the Hamiltonian (\ref{eq:H}) and the
perturbation (\ref{eq:V}) are real and symmetric. They have the
following forms ($n \ge m$):
\begin{align*}
&H^0_{m\, n} = \delta_{m \;n} \left[\hbar \omega \left(n + \frac12\right) +
\frac g 2 \left(\frac32 \, g\, a_0 \, (2 m^2 + 2  m + 1) - a'_2 (2 m + 1)
\right)  \right] \\  
&+ \delta_{m + 2 \; n} \;\frac{g}{2} \left(g\, a_0  (2 m + 3) - a'_2 \right)
\sqrt{(m + 1)(m  + 2)}\\
&+ \delta_{m + 4 \; n} \frac{a_0 g^2}{4} \sqrt{(m + 1)(m + 2)(m + 3)(m + 4)},\\
&x_{m\, n} = \delta_{m + 1 \; n} \;\sqrt{\frac g2} \; \sqrt{m + 1},
\end{align*}
where $g  = \hbar/m \omega$ and $a'_2 = a_2 + m \,\omega^2/2$, $\hbar$ is Planck
constant which we put equal to $1$, $\omega$ -
frequency of harmonic oscillator which is arbitrary, and so may be adjusted to
optimize the computation. We use the value $\omega =
0.2$ with parameters $m~=~1,$ $a_0~=~1/128,$ $a_2 = 1/4$. The matrix size is
chosen to be equal to $200 \times 200$. Calculations with larger matrices give
the same results. System of computer algebra Mathematica was used for numerical
calculations.

\begin{figure}[th]
\centering
\includegraphics[angle = 270,width=0.6\textwidth]{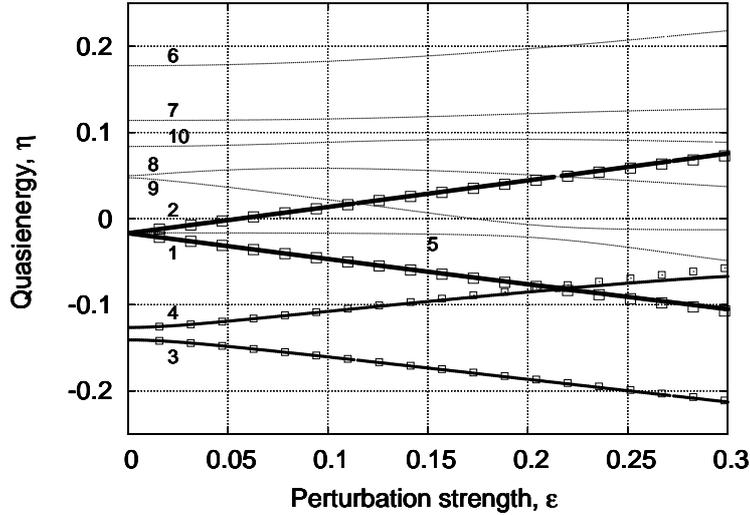}
\caption{Quasienergy spectrum for the ten lowest average energy levels. All
levels are numbered in order of the average energy values. Solid lines -
quasienergy levels for the kicked system. Thick lines - two doublets with the
minimal average energy. Empty squares - shifted energy levels of the asymmetric
model. The model parameters are $m~=~1,$ $a_0~=~1/128,$ $a_2 = 1/4$ and $\nu =
0.5$. }\label{fig:qeS}
\end{figure}

In order to obtain quasienergy levels ($\eta_k$) in the kicked double well
system directly we calculate eigenvalues ($\lambda_k$) of the one-period
evolution operator $e^{- i H T} e^{-i V}$ and express quasienergy levels through
the definition $\eta_k = i \, \ln \lambda_k/T$. Then we get ten levels with the
lowest one-period average energy which is calculated using the formula
$\left<v_i\right|H_0 + V/T\left|v_i\right>$ ($\left|v_i\right>$ are the
eigenvectors of the one-period evolution operator). The dependence of the
quasienergies of these ten levels on the strength of the perturbation is shown
in the figure~\ref{fig:qeS}. Quasienergies of the two doublets with the minimal
average energy (thick lines in the figure~\ref{fig:qeS}) has a linear dependence
on the strength of the perturbation in the considered parameter region. They are
strongly influenced by the perturbation while some of the quasienergy states are
not.

\begin{figure}[h!]
\centering
\includegraphics[angle = 270,width=0.48\textwidth]{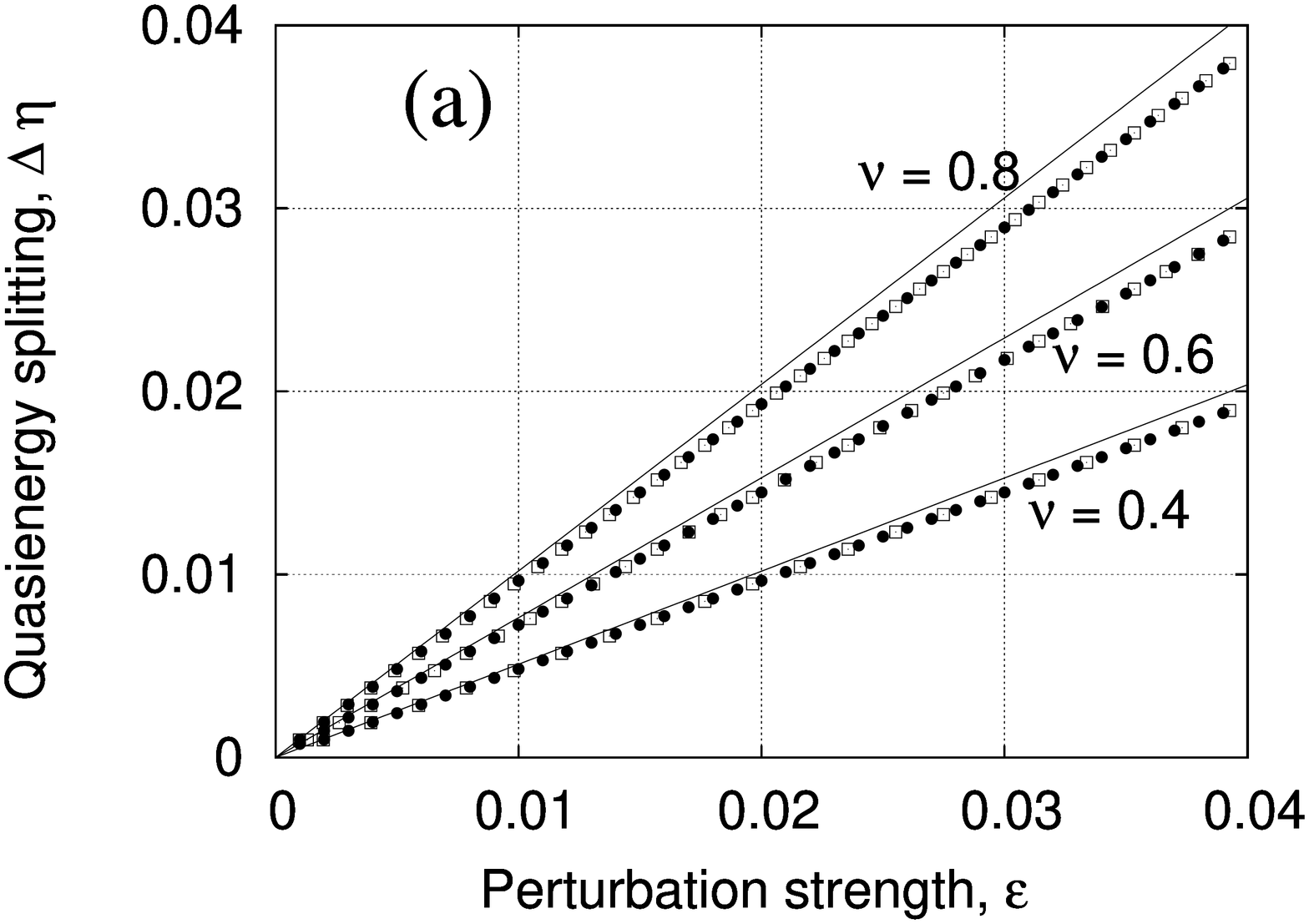}
\includegraphics[angle = 270,width=0.48\textwidth]{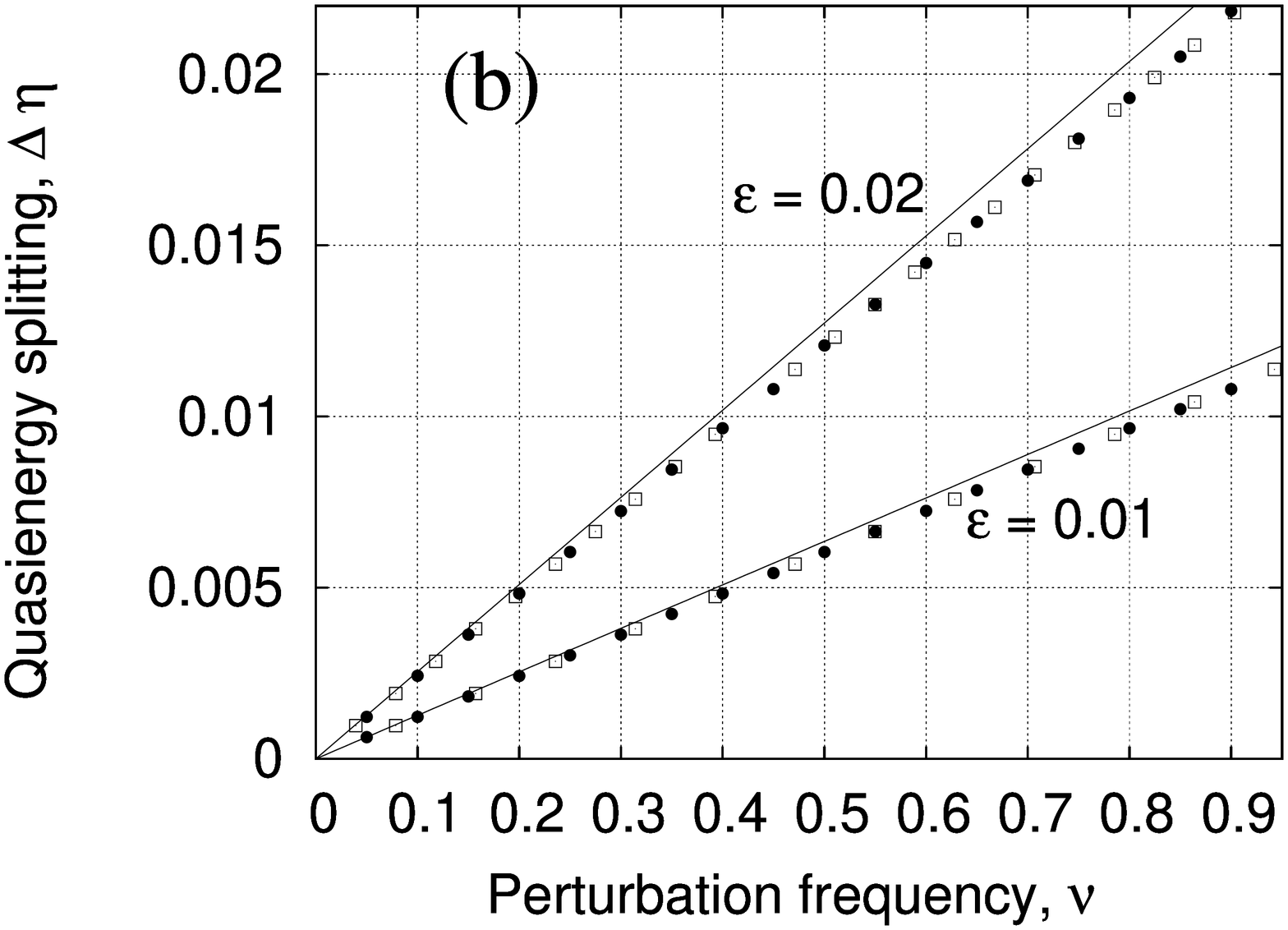}
\caption{Quasienergy splitting as a function of the strength~($a$) and
frequency~($b$) of the perturbation. Filled circles - results of the numerical
calculations for the kicked system. Empty squares - shifted energies for the
asymmetric model. The model parameters are $m~=~1,$ $a_0~=~1/128,$ $a_2 =
1/4$.}\label{fig:dE}
\end{figure}

To obtain the same levels in the framework of the effective model approach we
have to calculate the eigenvalues of the Hamiltonian of the asymmetric double
well potential (\ref{def:hamas}) with the asymmetric parameter defined by the
expression (\ref{def:sigma}). Then we shift obtained eigenvalues to zone
$(-\frac{\nu}{2},\frac{\nu}{2})$ in order to compare results with ones in the
kicked system. The result of calculations on the base of this procedure is shown
in the figure \ref{fig:qeS} by empty squares for two lowest doublets. Comparing
results of direct and model calculations we make the conclusion that the levels
of the doublets laying below the potential hump are correctly described by the
effective model. 

Performed numerical calculations for the kicked and the effective system give
the dependence of the ground quasienergy splitting both on the strength
(fig.\ref{fig:dE}(a)) and the frequency (fig.\ref{fig:dE}(b)) of the
perturbation. Filled circles in figures correspond to kicked double well system,
empty squares to numerical results obtained in the framework of the effective
model. There is good agreement between splitting's dependencies for these two
systems. They are linear as it was predicted by expression (\ref{an-fomula}). It
should be mentioned that all dependencies for the effective model was obtained
from one series of the numerical calculations. We fix model parameters $a_0,
a_2, m$ and calculate numerically one set of numerical points for the dependence
on the asymmetric parameter. In kicked system we should to perform one series of
numerical calculations for every dependency. This is the first advantage which
was discussed in the previous section. 

The second advantage which we should discuss after description of the used
numerical methods is a more simple algorithm of the calculations. In the kicked
system we should to calculate eigenvalues of the matrix exponents. This is more
difficult task than in asymmetric model where we calculate the eigenvalues of
the system Hamiltonian.

Analytical result (\ref{an-fomula}) which was put forward as third advantage of
the method is plotted in the figures~\ref{fig:dE}~(a) and \ref{fig:dE}~(b) by
straight solid lines. Numerical points lie close to these lines. The agreement
between numerical calculations and analytical expression (\ref{an-fomula}) is
good (near $6 \%$) in the parametric region considered.

\section{Conclusions}

Effective Hamiltonian for the kicked double well system was obtained using the
{Campbell}-{Baker}-{Hausdorff} expansion formula. Effective autonomous
asymmetric model for this system was constructed. This model is more convenient
in numerical calculations than kicked one. Results of numerical calculations
show that model correctly describes quasienergy spectrum of the kicked system
for low laying levels.

The analytical formula for the ground quasienergy splitting dependence on both
the perturbation strength and frequency was obtained in the framework of the
effective asymmetric model. This formula predicts linear dependence of the
ground  quasienergy splitting on these parameters for the small perturbation
strength and period values. Numerical results for the quasienergy splitting as a
function of the perturbation frequency and strength demonstrate linear
dependence as well. They are in a good agreement with the formula
(\ref{an-fomula}). Proposed approach will be used in future for investigation of
the CAT and CDT phenomena.

\end{document}